%% file: main.tex
\documentclass[sigconf]{acmart}

\usepackage{amssymb,makecell,pifont}
\usepackage{xcolor}
\usepackage{kotex}
\usepackage{tikz}
\usepackage{subcaption}
\usepackage{balance}
\AtBeginDocument{%
  }

\copyrightyear{2026}
\acmYear{2026}
\setcopyright{cc}
\setcctype{by}
\acmConference[WPES '26]{25th Workshop on Privacy in the Electronic Society}{November 15--19, 2026}{The Hague, Netherlands}
\acmBooktitle{25th Workshop on Privacy in the Electronic Society (WPES '26), November 15--19, 2026, The Hague, Netherlands}
\acmDOI{10.1145/3847192.3847361}
\acmISBN{979-8-4007-3026-9/2026/11}

\newcommand{\cmark}{\ding{51}}
\newcommand{\xmark}{\ding{55}} 
\newcommand{\ourtool}{\textsc{GAZEleak}}

\newcommand{\junhee}[1]{\textcolor{blue}{#1}}

\begin{document}

\title{\ourtool{}: Passcode Inference Against Eye-tracking XR Devices Through External Observation}

\author{Hwanjo Heo}
\authornote{Co-first authors.}
\email{hwanjo@etri.re.kr}
\orcid{0000-0002-8105-4224}
\affiliation{%
 \institution{ETRI}
 \city{Daejeon}
 \country{Republic of Korea}}

\author{Junhee Lee}
\authornotemark[1]
\email{junhee1004m@gmail.com}
\orcid{0009-0009-4967-3621}
\affiliation{%
 \institution{Chungbuk National University}
 \city{Cheongju}
 \country{Republic of Korea}}

\author{Jinwoo Kim}
\authornote{Corresponding author.}
\email{jinwookim@cbnu.ac.kr}
\orcid{0000-0003-1303-8668}
\affiliation{%
 \institution{Chungbuk National University}
 \city{Cheongju}
 \country{Republic of Korea}}

\newcommand{\cnum}[1]{%
  \tikz[baseline=(char.base)]{
    \node[shape=circle, fill=black, text=white,
          inner sep=1pt, minimum size=1.6ex,
          font=\footnotesize\bfseries] (char) {#1};
  }%
}



\input{sections/abstract}





\maketitle

\input{sections/introduction}

\input{sections/background-threat-model}
\input{sections/attack}
\input{sections/evaluation}

\input{sections/discussion}

\input{sections/related}
\input{sections/ack}
\input{sections/conclusion}

\bibliographystyle{ACM-Reference-Format}
\balance
\bibliography{reference}


\end{document}

%% file: sections/abstract.tex
\begin{abstract}
Mixed-reality headsets such as Apple Vision Pro replace the touchscreen with
\emph{gaze-as-pointer} interaction: the wearer looks at a target and confirms
with an air pinch. Because the display is inside the headset and the eye tracker
is walled off from third-party software, such input is widely assumed to be
unobservable to bystanders---a built-in defense against the shoulder-surfing
that plagues phones and laptops. We present \ourtool{},
a side-channel attack that recovers gaze-driven input from \emph{external video of head motion alone}, under a strictly local, physical-observer threat model:
the adversary only films the wearer from across the room and installs no software
on the device. The attack exploits the centrally coupled eye--head motor
program---gaze shifts recruit small, target-dependent head reorientations---which projects into sub-degree pose changes recoverable from
commodity video. 
\ourtool{} implements a measurement-based, sparse-optical-flow inference pipeline for users' 6-digit device passcode. 
On a preliminary front-view dataset from three author-subjects who were aware of the attack hypothesis, \ourtool{} places the true code within the top ten guesses for $56\%$ (10 of 18) of test codes under a cross-person protocol with no labeled victim data, and for every code of the most exposed subject. The performance is subject-dependent: no passcode from the least exposed subject reaches the top ten, although its median guessed passcode rank is 12{,}786 rather than 500{,}000 expected from an uninformative ordering. 
These results provide preliminary evidence that gaze-coupled head motion can expose passcode information under controlled conditions, while motivating broader evaluation across users, behaviors, and capture settings.
\end{abstract}

%% file: sections/introduction.tex
\section{Introduction}

Personal computing is moving beyond the smartphone toward wearable, head-mounted devices that promise always-on, hands-free interaction. The first commercial wave is already deployed: Apple Vision Pro~\cite{apple_vision_pro} and Meta Quest 3~\cite{meta_quest_3} ship as mixed-reality headsets, while smart glasses such as Meta Ray-Ban~\cite{rayban_meta} and the recently previewed Meta Orion~\cite{meta_orion} target an ambient-computing form factor. 
Apple Vision Pro exemplifies a shift in how users provide input:
the wearer gazes at an interface element to point and performs an air pinch to confirm the selection. Unlike a touchscreen tap or a handheld-controller movement, the pinch reveals when a selection occurs but not which virtual target was selected. 

\begin{figure}[t]
    \centering
    \includegraphics[width=\linewidth]{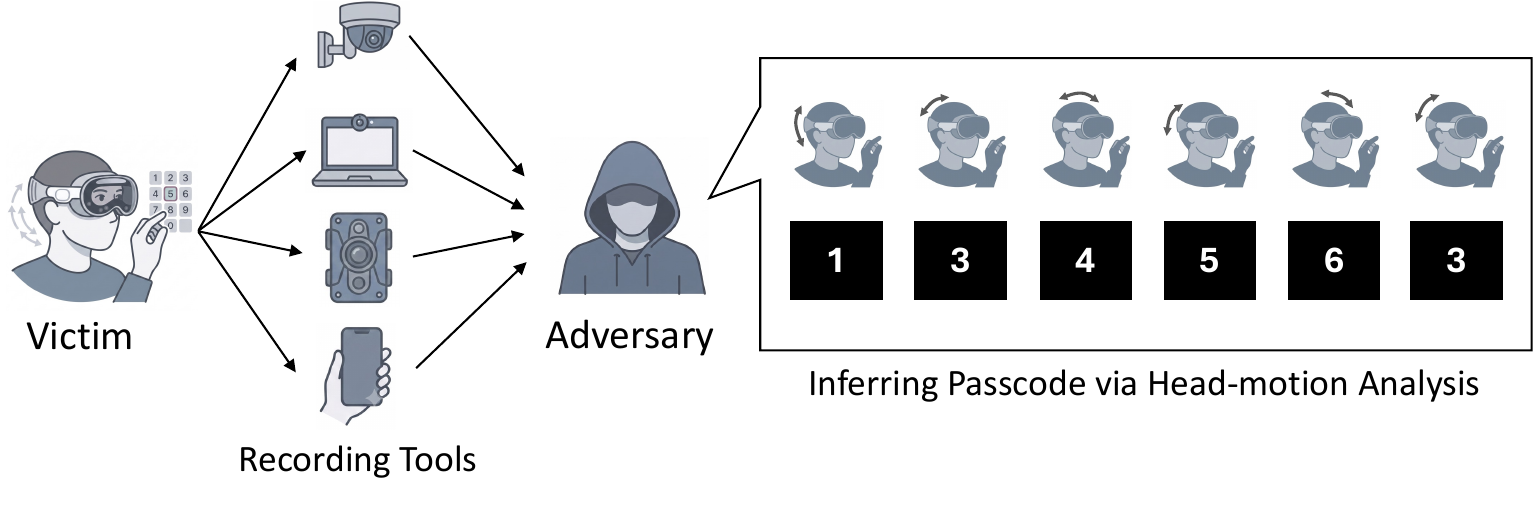}
    \caption{Our threat model: an adversary records the head motions of a victim wearing an XR device to infer their 6-digit passcode entered via air-pinch gestures.}
    \label{fig:threat_model}
\end{figure}

Even with this confirmation gesture visible, the interaction appears resistant to shoulder surfing for two reasons. First, the display is enclosed by the headset and is not directly visible from an ordinary external viewpoint. Second, the operating system restricts access to raw gaze data, preventing ordinary third-party applications from observing where the user looks. A nearby observer can therefore see the wearer and the confirmation gesture while remaining unable to see either the rendered user interface or the selected input. These protections conceal the \emph{digital} input channel, but they do not necessarily conceal the wearer's \emph{physical} response. Gaze shifts can recruit small, target-dependent head movements, and the headset makes those movements visible as image-plane translation. The selected target may consequently leave a physical side channel even when the display and eye-tracking stream remain inaccessible. \emph{The security question here is whether an adversary can infer sensitive information with some confidence by analyzing the victim's head motions recorded in external videos} (see Figure~\ref{fig:threat_model}).

To answer this question, we present \ourtool{}, a side-channel attack that infers gaze-entered passcodes using \emph{external video of an XR-headset wearer's head motion}.
The attack exploits the centrally-coupled eye--head motor program: gaze shifts can recruit small, target-dependent head movements~\cite{sidenmark2019eye}, and the resulting sub-degree pose changes are recoverable from commodity video using classical sparse feature tracking. While some XR keylogging attacks have been proposed so far, they do not address this physical adversary. TyPose~\cite{slocum2023going} inferred virtual-keyboard input from device-reported head-motion telemetry, while GAZEploit~\cite{wang2024gazeploit} recovered typed input from the animated eyes of a Vision Pro Persona transmitted over FaceTime. These attacks exploit signals exposed through the device or a software-rendered representation. We instead study a local and physical observer who records the victim directly and receives no cooperation from the headset, operating system, or third-party software. Device-side access control and avatar suppression do not remove the underlying motion signal, although interface changes can reduce its exploitability.

\noindent\textbf{Scope of this paper.} 
To evaluate \ourtool{}, we perform a controlled, preliminary feasibility study of six-digit gaze-and-pinch passcode entry only from frontal camera recordings of the author subjects. The results therefore establish signal availability under controlled conditions rather than a population-level estimate of attack success. A larger study with unaware participants, additional camera viewpoints, and input tasks beyond numeric passcodes is planned subject to institutional ethics approval. We discuss these limitations in Section~\ref{sec:limits}.

\noindent\textbf{Contributions.} We make the following contributions.
\begin{list}{\labelitemi}{\leftmargin=1em} 

    \item To our knowledge, we present the first passcode-inference attack
    against gaze-as-pointer XR interaction that relies solely on external video.
    The adversary requires no software on the target device and no access to its
    display, operating-system APIs, eye-tracking stream, or head-motion
    telemetry.

    \item We design and implement \ourtool{}, an end-to-end inference pipeline
    that detects visible air-pinch events, segments a passcode entry into
    inter-key transitions, measures sub-degree head motion using classical
    sparse feature tracking, predicts keypad displacements, and accumulates
    transition evidence to rank all $10^6$ six-digit passcode candidates.

    \item We conduct a controlled three-subject feasibility study under both
    within-person and cross-person protocols, using security-relevant outcomes
    that include top-10 accuracy and the true passcode's rank. Under the
    cross-person protocol, \ourtool{} places the true code within ten guesses
    for 56\% (10 of 18) of all test codes and for every code of the most exposed subject.

    \item We characterize the conditions that determine attack effectiveness
    through experiments on repeated observations, spatial and temporal capture
    degradation, transition-label granularity, scoring methods, descriptor
    selection, and vision model alternatives. 
\end{list}

The rest of the paper is organized as follows. Section~\ref{sec:background} provides necessary background and our motivation. Section~\ref{sec:threat} defines our adversary assumption. Section~\ref{sec:attack} and Section~\ref{sec:impl} present the design and implementation of \ourtool{}, respectively. Section~\ref{sec:evaluation} evaluates the attack under within- and cross-person protocols and examines the performance with various factors. Section~\ref{sec:discussion} discusses results and security implications. Section~\ref{sec:related} reviews related work, and Section~\ref{sec:conclusion} concludes the paper.

%% file: sections/background-threat-model.tex
\section{Background}
\label{sec:background}

This section provides the background and motivation for \ourtool{}. We first review eye tracking as an emerging input modality in consumer XR devices (Section~\ref{sec:bg-eye-track}), then describe the eye–head coordination that makes gaze shifts externally observable (Section~\ref{sec:bg-eye-head}), and finally explain how this physical coupling motivates our attack (Section~\ref{sec:motivation}).

\subsection{Eye-Tracking Capability in XR Devices}
\label{sec:bg-eye-track}

Eye tracking has shifted from a laboratory peripheral to a built-in
sensor in mainstream consumer headsets. The Apple Vision Pro
is the most notable example: rather than treating gaze as an
auxiliary signal for foveated rendering or social-presence avatars,
Vision Pro elevates it to the \emph{primary pointing modality}, with
the operating system inferring the user's target object from where
the eye is looking and confirming selection through an air pinch.
Comparable hardware exists on the Meta Quest Pro~\cite{meta_quest_pro_eyetracking}, which
integrates Tobii-based gaze trackers behind the lenses, and HTC's
Vive Pro Eye and Vive XR Elite~\cite{htc_vive}, although these primarily use
gaze for rendering optimization or social-VR animation rather than as
a UI pointer. Several smaller platforms (e.g., Pico Neo 3 Pro Eye, Pico 4
Pro, PlayStation VR2) ship with eye-tracking subsystems as well.

Eye tracking has not yet become standard in lighter \emph{AI glasses} form factor. 
Meta's commercial Ray-Ban smart glasses include cameras and audio but no
eye-tracking sensor. The recently-previewed Meta Orion prototype and
the research-oriented Meta Aria~Gen~2 platform~\cite{rice2025egodrive} both
incorporate inward-facing eye cameras, signaling that mainstream AI
glasses are converging toward the same gaze-as-input paradigm
established by current XR headsets. We therefore treat gaze-as-pointer
as the de~facto interaction modality across the XR
device class our threat model targets.

\subsection{Synchronization of Eye and Head Movement}
\label{sec:bg-eye-head}

A long-standing finding in motor neuroscience is that eye and head
movements are not independently controlled but driven by a single
coordinated motor program. Sidenmark and
Gellersen~\cite{sidenmark2019eye} establish this directly in a VR
setting: across 7{,}600 gaze shifts collected from 20 participants,
they report that natural gaze re-orientation systematically recruits
all three of the eyes, head, and torso, and that the relative
contribution of each component varies as a function of the target's
angular position. Earlier work in non-immersive settings reaches the
same conclusion --- gaze shifts decompose into an eye component and a
head component driven by a centrally coordinated command~\cite{bizzi1971eye}. 
In the meantime, it is also reported that the head barely contributes 
when targets fall within a comfortable oculomotor range, 
but as the angular demand grows the head increasingly compensates~\cite{stahl1999amplitude}.

Crucially, the eye--head coupling is largely involuntary and cannot simply be suppressed at will.
Corneil et al.~\cite{corneil2004visual} record neck-muscle EMG
in head-restrained primates and observe that the appearance of a
visual target evokes a sharply time-locked, lateralized activation
of the neck muscles within roughly 55--95~ms of stimulus onset, even
though the head itself is physically prevented from moving. The
central motor command for head rotation is therefore issued in
parallel with, and often preceding, the eye saccade; mechanical restraint hides the overt
movement but does not prevent its generation. In an XR setting, where
the wearer is unrestrained, the small but consistent head
motion accompanying each gaze shift may therefore be observable to an
external camera that can resolve sub-degree head pose --- the side
channel that underlies the attack we present in this paper.

\subsection{Motivation}
\label{sec:motivation}
Our study is motivated by direct observation of how people use gaze-as-pointer XR devices in practice. While working alongside colleagues who routinely use headsets such as Apple Vision Pro, one of the authors noticed that most users do not hold the head still and move only the eyes. Instead, they reorient the head noticeably toward on-screen targets, to the point that the movement is plainly visible to a bystander. This is the everyday manifestation of the eye--head coupling reviewed in Section~\ref{sec:bg-eye-head}: gaze-recruited head motion is not a corner case but the common mode of interaction, and it is overt enough to be captured by an ordinary camera.

The tendency appears especially pronounced during authentication processes. When users unlock the headset with a passcode or type a password on a virtual keyboard, they tend to enter each character deliberately, orienting the head toward the target key and exaggerating the movement relative to casual browsing.\footnote{A similar observation motivates a VR keylogging attack that requires malicious software to be installed on the headset~\cite{slocum2023going}.} An informal conversation with several users suggested why: during authentication they are anxious about input errors, because a mistaken digit or character must be undone with awkward gaze-driven backspacing, so they slow down and aim carefully at each key. The care that improves their entry accuracy also produces larger, cleaner, more target-specific head movements.

This observation is what makes passcode entry a compelling target rather than a hypothetical one. The setting in which the user is most careful---security-sensitive authentication---is precisely the setting in which the \emph{gaze-coupled head-motion side channel} is strongest. We note that this accuracy-driven exaggeration is distinct from, and works against, the behavior of an attack-aware user who might consciously suppress head motion (which we treat as a limitation in Section~\ref{sec:limits}): a typical user, unaware of the attack and focused only on entering the code correctly, is inclined to move more. These observations are informal and motivate our evaluation with real measurements and also the systematic user study we plan as future work.

\section{Threat Model}
\label{sec:threat}

Figure~\ref{fig:threat_model} illustrates our threat model. We assume that a victim is a user of a commercial XR headset or AI-glasses whose primary or co-primary pointing modality is eye gaze. Our running example is Apple Vision Pro, in which the OS recognizes the target object from the user's gaze and the user confirms with an air pinch. The model generalizes to other eye-tracked devices such as Meta Quest Pro, HTC Vive Pro Eye, and the emerging class of AI glasses with inward-facing eye cameras.

The adversary is strictly \emph{external}. They do not install or compromise software on the device, do not query any OS-level API, and do not have access to the inward micro-display, the eye-tracker stream, or the on-device gaze coordinates. They cannot mount any software-level side-channel attack on the headset. The adversary can capture video of the victim's head and upper body from a passive vantage point: a smartphone held discreetly in the same room, a fixed surveillance camera, a webcam on a nearby laptop, or a body-worn camera. Recordings are short (seconds to a minute) at standard consumer resolution and frame rate (e.g., 4K/30fps), with no calibration to the victim or the device. 

The adversary aims to exfiltrate sensitive gaze-driven input by the victim. Specifically, in this paper, we aim to infer the six-digit passcode that is entered with gaze selection and air-pinch confirmation. 
The goal of the adversary is to infer the correct six-digit passcode from $10^6$ possible candidates. In many practical passcode
settings, several retries may be possible before lockout or escalation, and an
attacker who can reduce the search space to a top-$k$ list has substantially
increased risk. We choose $k=10$ as the Apple Vision Pro allows up to 10 failed passcode attempts~\cite{vision_pro_passcode}.

Among a number of prospective attack scenarios by an adversary with gaze-inferring capability, we consider the passcode inference by an adversary with opportunistic passcode input session capture. The victim wakes their XR headset after sleep, a power cycle, or an inactivity timeout and is prompted for a 6-digit device passcode.\footnote{Apple Vision Pro requires the user to input the passcode, even when Optic ID (Apple's iris-based biometric authentication system) is enabled, after a reboot or an extended period of inactivity for security purposes~\cite{apple-optic-id}.} The adversary is co-located (a coffee shop, an office, a lobby) and records the brief ($<10$~s) entry from a smartphone in hand or pocket, or from a CCTV camera overhead. A single capture may be noisy, but the adversary may be able to accumulate multiple recordings of the same passcode over time.

%% file: sections/attack.tex
\section{\ourtool{} Design}
\label{sec:attack}

This section describes the design of \ourtool{}. We first give an overview of the whole attack pipeline and formalize the passcode as a sequence of keypad transitions (Section~\ref{sec:transition-model}). The following subsections then detail each inference stage, from transition segmentation (Section~\ref{subsec:segmentation}) and motion description (Section~\ref{subsec:description}) to label scoring (Section~\ref{sec:scoring}) and passcode ranking (Section~\ref{subsec:ranking}).

\begin{figure*}[t]
    \centering
    \includegraphics[width=\linewidth]{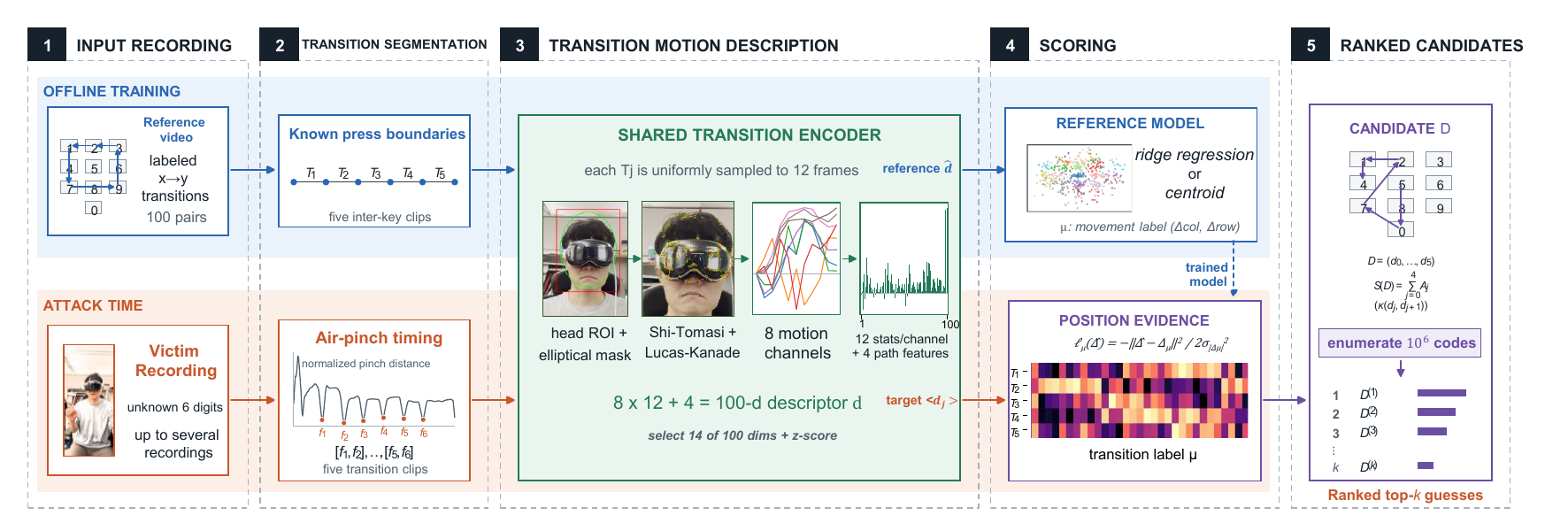}
    \caption{Overview of \ourtool{} attack pipeline for 6-digit passcode inference.}
    
    \label{fig:overview}
\end{figure*}

\subsection{Overview}

Figure~\ref{fig:overview} illustrates the passcode inference pipeline of \ourtool{},
which converts an external passcode-entry video into a ranked list of six-digit
candidates. End-to-end, \ourtool{} \cnum{1} detects six air-pinch events from the victim
recording, \cnum{2} segments the five intervening head-motion transitions
(Section~\ref{subsec:segmentation}), \cnum{3} encodes each transition into a fixed-length
motion descriptor (Section~\ref{subsec:description}), \cnum{4} maps each descriptor to a
keypad displacement and scores it against every candidate transition label
(Section~\ref{sec:scoring}), and \cnum{5} pools the per-position evidence to rank all
$10^6$ candidate passcodes (Section~\ref{subsec:ranking}). The unknown passcode is
not used during feature extraction, scoring, or ranking. In addition, \ourtool{} operates in two modes: (i) offline training where it learns transitions from a victim or others, and (ii) attack time that infers the passcode of a victim.

\subsection{Transition-Based Passcode Model}
\label{sec:transition-model}

To formalize the gaze-inference problem, we model the passcode as a sequence of \emph{transitions} rather than as six
independent digit observations. Let
$D=(D_0,\ldots,D_5)$ be a six-digit passcode. Here, we define three transition labels at different granularities. The \textbf{exact-pair} label at
position $j$ is
\[
  \pi(D_j,D_{j+1}) = 10 \cdot D_j + D_{j+1},
  \qquad j\in\{0,\ldots,4\}.
\]
Thus, the numeric keypad induces 100 exact transition labels.
For robustness analysis, we also consider coarser label spaces. Let each digit
have a keypad coordinate $(\mathrm{col}(D_j),\mathrm{row}(D_j))$.
The \textbf{movement} label is:
\[
  \mu(D_j,D_{j+1})=
  \bigl(\Delta\mathrm{col},
        \Delta\mathrm{row}\bigr),
\]
which groups exact pairs that share the same keypad displacement where $\Delta\mathrm{col} = \mathrm{col}(D_{j+1})-\mathrm{col}(D_j)$ and $\Delta\mathrm{row} = \mathrm{row}(D_{j+1})-\mathrm{row}(D_j)$. The \textbf{direction}
label further coarsens this to:
\[
  \delta(D_j,D_{j+1})=
  \bigl(\mathrm{sgn}(\Delta\mathrm{col}),
        \mathrm{sgn}(\Delta\mathrm{row})\bigr).
\]
At the two extremes, exact-pair labels retain absolute source and destination information, whereas direction labels retain only the signs of keypad displacement. Consequently, direction scoring constrains the passcode path but generally leaves multiple exact passcodes indistinguishable unless additional absolute-position evidence is available. \ourtool{} adopts the movement label $\mu$ as its readout (Section~\ref{sec:scoring}); we treat exact-pair and direction as alternatives evaluated in the ablation (Section~\ref{sec:ablation}).

\subsection{Transition Segmentation}
\label{subsec:segmentation}

As each digit selection is confirmed by an air pinch motion, the six presses partition the recording into five transitions. Because the keypad is invisible to the camera, we recover the press timing from the wearer's hand alone. For every frame, we run a single-hand landmark detector~\cite{zhang2020mediapipe} and compute the
thumb-tip--to--index-tip distance, normalized by an intrinsic hand scale (wrist to index knuckle) so that it is invariant to hand size and camera distance. A pinch is a sharp local minimum of this distance (the fingertips touch); we smooth the signal, extract local minima, and take the six deepest, subject to a minimum temporal separation, as the press frames $f_0<\cdots<f_5$. The five transition clips are the intervals $[f_0,f_1],[f_1,f_2],\dots,[f_4,f_5]$. For labeled transition video clips used at the training phase (Section~\ref{sec:evaluation}), CSV recorded key input boundaries are utilized for convenience.   

\subsection{Transition Motion Description}
\label{subsec:description}

The signal is measured inside a per-view head/headset region of interest (ROI) $B=(x_0,y_0,x_1,y_1)$, chosen to contain stable headset, visor, face, and head-boundary texture while excluding the hands and as much background as possible. We localize $B$ automatically with a promptable segmentation model (e.g., SAM~2~\cite{ravi2025sam}) and confirm it manually.
The resulting ROI is fixed within a recording, but separate recordings may use different boxes.

Within the ROI, we detect feature points only inside an inscribed elliptical mask. The head is approximately elliptical, so the rectangle corners are the pixels most likely to hold static background; the ellipse suppresses those background and ROI-corner points while retaining the high-gradient head boundary. The mask constrains only the initial point detection---its boundary is never used as a motion cue---which avoids segmentation jitter while preserving stable texture for optical flow.

We encode each transition clip with a fixed-length descriptor obtained from sparse optical flow, following the classical detect-then-track paradigm. Each clip is temporally resampled to $N=12$ frames (uniform indices over $[f_k,f_{k+1}]$) so descriptors are invariant to transition duration.
For each transition clip, we crop the ROI from the sampled frames and convert the crops to grayscale. On the first frame we detect Shi--Tomasi corners~\cite{shi1994good}. The selected points are then
tracked across the sampled frames with pyramidal Lucas--Kanade optical
flow~\cite{bouguet2001pyramidal}. Points that fail the
tracker status check are dropped.

Let $\mathbf{p}^{(0)}_j=(x^{(0)}_j,y^{(0)}_j)^\top$ be the location of tracked
corner $j$ in the first sampled frame $f_0$ and let
$\mathbf{p}^{(i)}_j=(x^{(i)}_j,y^{(i)}_j)^\top$ be its Lucas--Kanade location in
the $i$-th sampled frame $f_i$. From these correspondences, we estimate the dominant 2-D
similarity transform with RANSAC~\cite{fischler1981random}:
\begin{equation*}
\label{eq:similarity-transform}
 \widehat{\mathbf{p}}^{(i)}_j=
 \begin{bmatrix}
  a_i & -b_i & t_{x,i} \\
  b_i &  a_i & t_{y,i}
 \end{bmatrix}
 \begin{bmatrix}x^{(0)}_j & y^{(0)}_j & 1\end{bmatrix}^{\!\top}.
\end{equation*}
Here, $(t_{x,i},t_{y,i})$ is translation in ROI pixels, while $a_i$ and $b_i$
encode isotropic scale $\rho_i=\sqrt{a_i^2+b_i^2}$ and in-plane rotation
$\theta_i=\operatorname{atan2}(b_i,a_i)$. The constrained model excludes shear
and nonuniform scale, which are not expected from short head movements, and
RANSAC suppresses static-background features and tracking outliers that remain
inside the ROI. We normalize translation as $(t_{x,i}/W,t_{y,i}/H)$ using the
ROI width $W$ and height $H$.

Because a single fitted transform can be unstable when few tracks agree, we
also retain robust statistics of the raw point flow
$\mathbf{u}_{j,i}=\mathbf{p}^{(i)}_j-\mathbf{p}^{(0)}_j$. The per-frame motion
vector is
\begin{equation*}
\label{eq:per-frame-motion-vector}
\resizebox{\columnwidth}{!}{$
 \mathbf{m}_i=\left(
 \frac{t_{x,i}}{W},\frac{t_{y,i}}{H},\theta_i,\rho_i-1,
 \frac{\operatorname{med}_j u_{x,j,i}}{W},
 \frac{\operatorname{med}_j u_{y,j,i}}{H},
 \frac{\operatorname{IQR}_j u_{x,j,i}}{W},
 \frac{\operatorname{IQR}_j u_{y,j,i}}{H}
 \right),
 $}
\end{equation*}
comprising transform translation, rotation, scale change, median point flow,
and point-flow dispersion. The sequence
$\mathbf{M}=(\mathbf{m}_1,\ldots,\mathbf{m}_{N-1})$ therefore contains eight
motion channels over the $N-1$ frame-to-reference estimates.

To obtain a duration-invariant descriptor we summarize each channel
over the series with twelve statistics capturing both its endpoint and its
dynamics (final value, mean, standard deviation, etc). We append four path-geometry scalars computed from the
translation and rotation channels---net displacement magnitude, total translation
path length, final rotation angle, and total absolute angular variation---that
describe the trajectory shape. This gives
a $8\times 12 + 4 = 100$-dimensional transition descriptor
$\mathbf{d}\in\mathbb{R}^{100}$.

\subsection{Label Scoring}
\label{sec:scoring}
 
We extract a fixed 14-dimensional subset of the 100-dimensional descriptor, chosen for stable head-motion features, and normalize each kept dimension with a z-score $\hat{x}_j=(x_j-m_j)/s_j$.\footnote{The descriptor-subset choice, i.e., retaining 14 of the total of 100 dimensions, is validated by our ablation study in Section~\ref{sec:subset}.} The mean $m_j$ and the standard deviation $s_j$ come from the reference set when the target provides labeled transitions, and from the target's own transitions otherwise. In the cross-person protocol, the adversary has no labeled data from the victim but does have the recordings of the attacked passcode (five entries; Section~\ref{subsec:repeat}). We therefore compute $m_j$ and $s_j$ from the transitions of those recordings. This removes the victim's constant head-pose bias before the regressor is applied.

For each transition, a ridge regressor predicts the keypad movement
$\hat{\mu}=(\widehat{\Delta\mathrm{col}},\widehat{\Delta\mathrm{row}})$ from the
standardized descriptor. 
It is trained on the reference transitions, whose movements are known. In the cross-person protocol (Section~\ref{subsec:settings}) the target provides no labels, so we refine the pooled regressor by unsupervised self-calibration: we repeatedly snap each predicted movement to the nearest valid keypad move, refit the map to those assignments, and iterate. This aligns the reference map to the target's own motion scale without using any target label.

We score each candidate movement $\mu$ by how close it is to the prediction
$\hat{\mu}$, with a Gaussian on their gap, written
$\ell(\mu)=-\lVert\hat{\mu}-\mu\rVert^2/(2\sigma^2_{|\mu|})$, where the variance
grows with the movement magnitude so that large movements are judged with a
wider tolerance. A softmax over labels turns these scores into a per-position
log-posterior,
\[
  \log P(\mu\mid\hat{\mathbf{x}})=\ell(\mu)-\log\!\sum_{\mu'} e^{\ell(\mu')},
\]
which the pooling step below accumulates.

\subsection{Passcode Ranking by Position Pooling}
\label{subsec:ranking}

Each target transition belongs
to one of the five transition indices $j\in\{0,\ldots,4\}$ within the passcode sequence. We
accumulate the per-position label evidence by summing the log-posteriors of the
transitions observed at that position,
$A_j(\mu)=\sum_{i:\,\mathrm{pos}(i)=j}\log P(\mu \mid\hat{\mathbf{x}}_i)$, and floor
labels unseen in the reference to a small log-probability. A candidate passcode
$D=(D_0,\dots,D_5)$ is then scored by summing the position-appropriate label
evidence over its five transitions,
$$
  S(D)=\sum_{j=0}^{4} A_j\bigl(\mu(D_j,D_{j+1})\bigr).
$$
We enumerate all $10^{6}$ candidate codes, sort by $S(D)$ in descending order, and
report the rank of the true code---equivalently, the attacker's ordered guess
list, of which the top~$k$ matters when the device permits $k$ attempts
(Section~\ref{sec:threat}).

%% file: sections/evaluation.tex
\section{Implementation}
\label{sec:impl}
We implemented the full pipeline of \ourtool{} and report the concrete implementation details here for reproducibility. 

\begin{figure}[t]
    \centering
    \centering
    \includegraphics[width=0.51\linewidth]{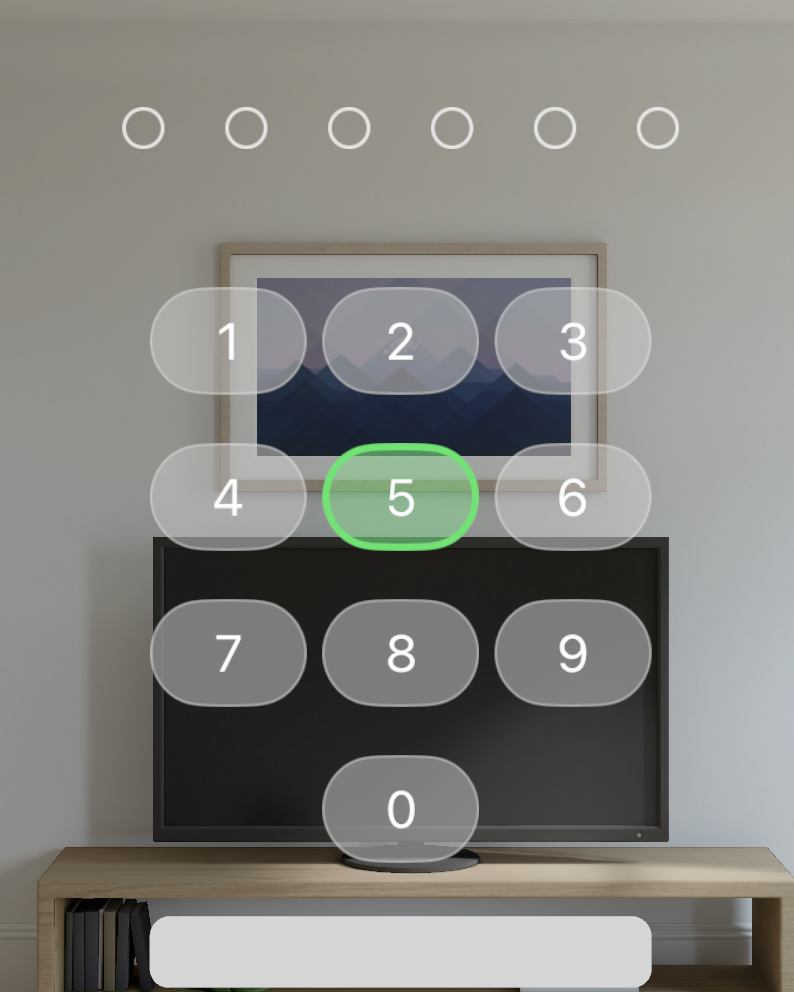}
    \hfill
    \includegraphics[width=0.48\linewidth]
    {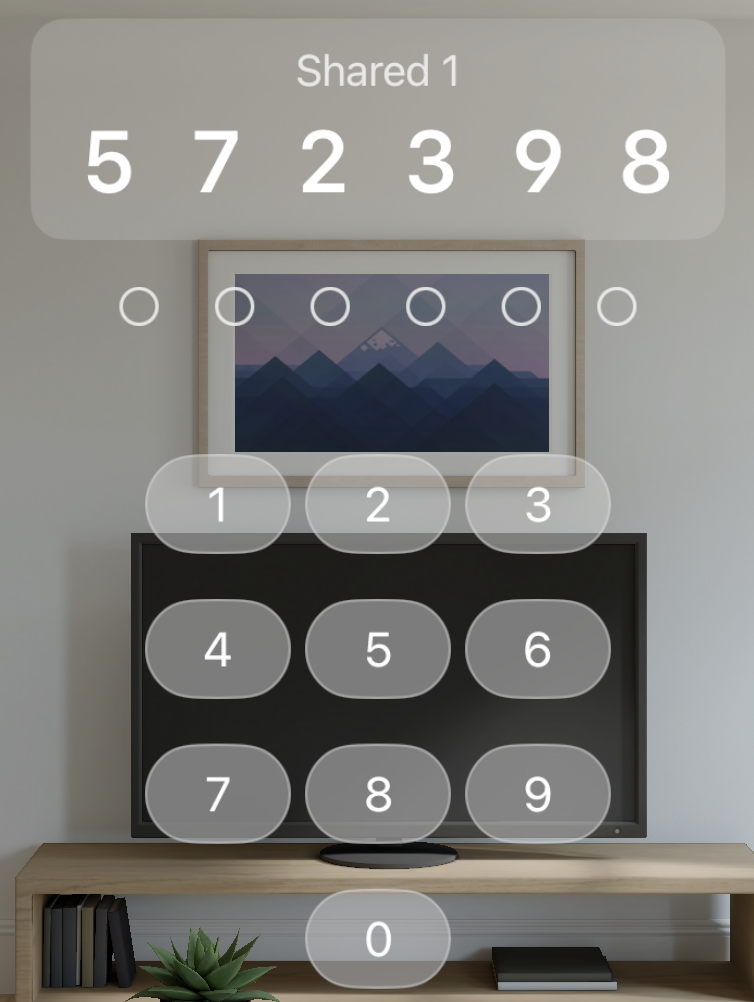}
    \caption{VisionOS app's keypad layout mimics the system passcode interface. Participants are instructed to input the highlighted (green) digit for training sessions (left) or assigned 6-digit passcode for test sessions (right).}
    \label{fig:keypad_image}
\end{figure}

\noindent\textbf{\ourtool{} VisionOS App.}  
To collect labeled data with precise ground truth, we built a data-collection
application\footnote{%
Our app is a measurement instrument for constructing our dataset;
it is \emph{not} part of the attack. The adversary in our threat model
(Section~\ref{sec:threat}) installs no software on the victim's device and uses
only external video.} for Apple Vision Pro that reproduces the system numeric passcode
keypad---a $3\times4$ grid whose key size and placement match the native
passcode screen---rendered at a fixed position in the immersive scene (Figure~\ref{fig:keypad_image}). The user selects keys with the device's native
gaze pointer and air-pinch confirmation, so the captured behavior matches
ordinary gaze-as-pointer input. The app was built with Swift 5.1
for visionOS 2.5.

The app runs two session types. \emph{Training} sessions prompt the user with
target digits and log each accepted press: a short Phase~1 warm-up that visits
the digits $0$--$9$ in order, followed by Phase~2, which traverses the ordered digit-to-digit transitions in a Eulerian sequence and repeats that traversal until the session reaches roughly $300$ transitions, so every transition is collected the same number of times.
\emph{Test} sessions prompt the user to enter six six-digit passcodes---three \emph{Shared} codes that are identical across profiles and
three \emph{Personal} codes unique to each profile---and each press is logged with its profile,
passcode index, position within the code, and entered digit.
The visionOS app streams these events over the local network to a companion macOS app, which stores them in CSV format.

\noindent\textbf{\ourtool{} Training/Inference Pipeline.}
All training and inference stages run offline on the recorded video; no component runs on or communicates with the headset. Recordings are made with a single commodity smartphone camera placed in front of the seated wearer and encoded at 4K/30fps. Frames are processed at native resolution; no per-wearer or per-device calibration is applied.
The attack pipeline is implemented offline in Python with OpenCV and it runs on Intel Xeon Gold 6526Y Linux server with NVIDIA H100 and no component executes on the headset.

Each transition clip is resampled to $N=12$ frames. Inside the elliptical ROI mask we detect up to $450$ Shi--Tomasi corners (quality level $0.004$, minimum spacing $7$~px) and track them with pyramidal Lucas--Kanade, then fit a partial-affine transform with RANSAC to suppress background and outlier tracks. Each transition yields a $100$-dimensional descriptor, of which a fixed $14$-dimensional subset is standardized and mapped to a keypad displacement by a ridge regressor with regularization $\lambda=1.0$.

\section{Evaluation}
\label{sec:evaluation}

This section evaluates \ourtool{}. In particular, we aim to answer the following research questions (RQ):

\begin{itemize}
    \item \textbf{RQ1.} Does \ourtool{} accurately infer the victim's transitions and passcode? (Section~\ref{sec:inference-results})
    \item \textbf{RQ2.} Does observing multiple entries of the same passcode improve inference? (Section~\ref{subsec:repeat})
    \item \textbf{RQ3.}  Does passcode difficulty affect inference performance? (Section~\ref{subsec:codediff})
    \item \textbf{RQ4.} Does the quality of the captured video affect inference performance? (Section~\ref{subsec:quality})
    \item \textbf{RQ5.} Which design components of \ourtool{} contribute to the inference performance? (Section~\ref{sec:ablation})
\end{itemize}

\subsection{Experiment Settings}
\label{subsec:settings}

\textbf{Data Collection.} We collected a controlled dataset of passcode-style gaze input on an Apple Vision Pro, as a representative XR platform. A participant wore the headset and entered digits on the numeric keypad, whose size and position were matched to the Apple Vision Pro system layout (Figure~\ref{fig:keypad_image}), while a single RGB camera recorded the wearer head-on in 4K resolution. Each session follows an Eulerian sequence that visits every ordered digit-to-digit transition uniformly, so that all pairwise head movements between the ten keys are represented, and none is over-sampled; this reference session supplies the labeled transitions used to fit the scorer. Separate held-out sessions provide six-digit test passcodes.

\noindent\textbf{Participants.} All three authors served as subjects. As no external participants were involved, this study did not require institutional review (Section~\ref{sec:ethics}). We report each subject separately rather than averaging, because the amount of head motion a wearer produces differs from person to person and that difference turns out to drive the result.

\noindent\textbf{Metrics.} To measure how \ourtool{} infers the victim's transitions, we measure \textit{transition accuracy}, the fraction of the five transitions per code whose predicted keypad displacement lands on the correct destination key, for which chance is $0.10$. To see how \ourtool{} infers the victim's passcode, we use \textit{top-10 accuracy}, the fraction of test codes whose true code appears within the attacker's first ten ranked guesses, which matches the ten failed attempts that Apple Vision Pro permits before lockout (Section~\ref{sec:threat}). Note that a uniform guesser reaches $10/10^6$. Additionally, we use \textit{median rank}, the median position of the true code when all $10^6$ candidates are sorted by the attacker's predicted likelihood, where a uniform guesser sits at $500{,}000$. Passcodes with identical movement sequences receive identical scores; none of the 18 test passcodes shares its movement sequence with another valid passcode, so ties did not affect the reported ranks.

\noindent\textbf{Protocols.} We frame each evaluation as an attack on one \emph{victim},
and report two protocols that differ in where the attacker's reference
transitions come from. In the \emph{within-person} protocol, the training
session and the test codes come from the same person, so the victim supplies
their own labeled transitions. In the \emph{cross-person} protocol, the
reference transitions come from the other two subjects and the victim
contributes no labels; an unsupervised self-calibration step aligns the pooled
map to the victim's own transition statistics. With only two reference subjects, this protocol measures transfer from a small reference pool rather than population-level generalization.

\subsection{Passcode Recovery Performance}
\label{sec:inference-results}

Table~\ref{tab:per-victim} reports transition accuracy and passcode recovery results. Performance varies substantially across the three subjects. For Subject 1, every test passcode appears within the top ten under both protocols, with median rank 1 within-person and 3 cross-person. Subject 2 reaches top-10 accuracy of 0.50 within-person and 0.67 cross-person, with median ranks 9 and 6, respectively. Subject 3 has no top-10 recovery: its transition accuracy is 0.30 within-person and 0.27 cross-person, and its median passcode ranks are 9,528 and 12,786. Across the three subjects, the cross-person protocol recovers 10 of the 18 test passcodes (56\%) within ten guesses, and the within-person protocol recovers 9 of 18 (50\%). Although Subject 3 does not meet the operational top-10 criterion (Section~\ref{sec:threat}), its ranking remains more informative than an unstructured search. We discuss the operational implications in Section~\ref{sec:discuss:search-space-impact}.

\begin{table}[!t]
\centering
\caption{Per-subject inference performance under the within- and cross-person protocols. \emph{Transition accuracy} is the fraction of the five inter-digit transitions in each six-digit passcode whose predicted displacement reaches the correct destination key (chance: 0.10). \emph{Top-10 accuracy} is the fraction of test passcodes whose true code is ranked among the first ten of $10^6$ candidates, with the number of recovered codes out of the six test passcodes per subject in parentheses. \emph{Median rank} is the median position of the true code in the ranked list (lower is better; an uninformative ordering has median rank 500,000). Every result pools the five recordings of each test passcode ($N=5$). Section~\ref{subsec:repeat} varies the observation budget.}

\label{tab:per-victim}
\small
\resizebox{\columnwidth}{!}{%
\begin{tabular}{l c c c c c c}
\toprule
& \multicolumn{3}{c}{\textbf{Within-person}} & \multicolumn{3}{c}{\textbf{Cross-person}} \\
\cmidrule(lr){2-4}\cmidrule(lr){5-7}
& \textbf{Transition} & \multicolumn{2}{c}{\textbf{Passcode}} & \textbf{Transition} & \multicolumn{2}{c}{\textbf{Passcode}} \\
\cmidrule(lr){2-2}\cmidrule(lr){3-4}\cmidrule(lr){5-5}\cmidrule(lr){6-7}
\textbf{Subject} & \textbf{Acc.} & \makecell{\textbf{Top-10}\\\textbf{Acc.}} & \makecell{\textbf{Med.}\\\textbf{Rank}}
                 & \textbf{Acc.} & \makecell{\textbf{Top-10}\\\textbf{Acc.}} & \makecell{\textbf{Med.}\\\textbf{Rank}} \\
\midrule
Subject 1 & 0.93 & 1.00 (6/6) & 1        & 0.80 & 1.00 (6/6) & 3 \\
Subject 2 & 0.73 & 0.50 (3/6) & 9        & 0.67 & 0.67 (4/6) & 6 \\
Subject 3 & 0.30 & 0.00 (0/6) & 9{,}528  & 0.27 & 0.00 (0/6) & 12{,}786 \\
\bottomrule
\end{tabular}%
}
\end{table}

\begin{figure}[t]
    \centering
    \includegraphics[width=\linewidth]{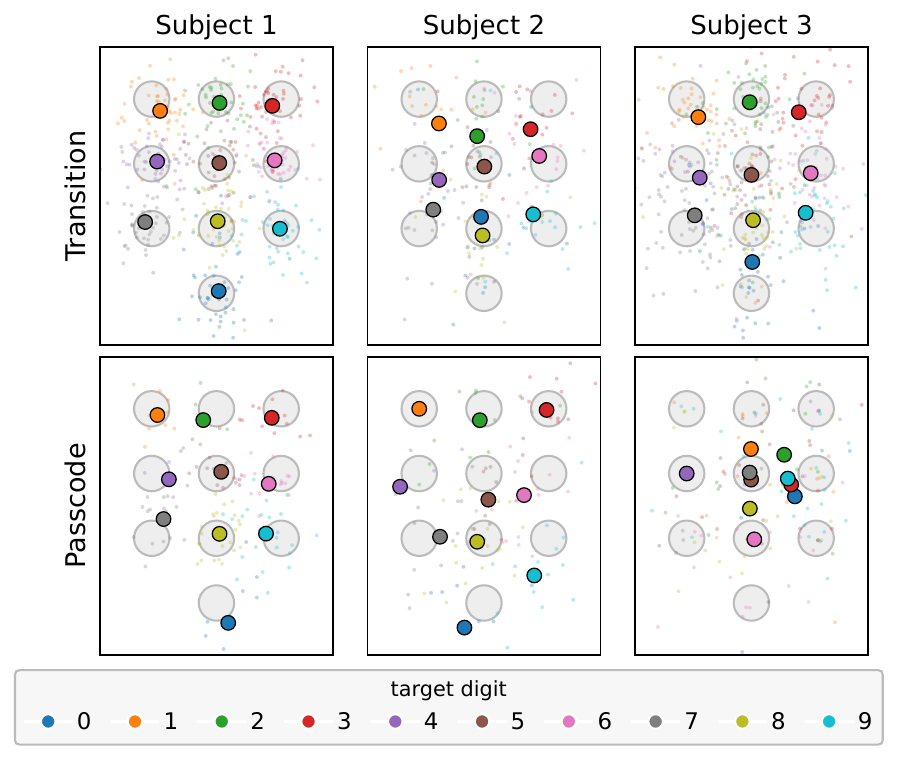}
    \caption{Recovered keypad geometry from (top) labeled transition and (bottom) passcode entry sessions. Each small point is a predicted destination position, colored by its ground-truth destination digit for evaluation; the larger colored markers show each digit's recovered keypad position, taken as the centroid of its predicted landing points, and the gray circles mark the nominal keypad location.}
    \label{fig:calib-vs-passcode}
\end{figure}

\begin{figure}[t]
    \centering
    \includegraphics[width=\linewidth]{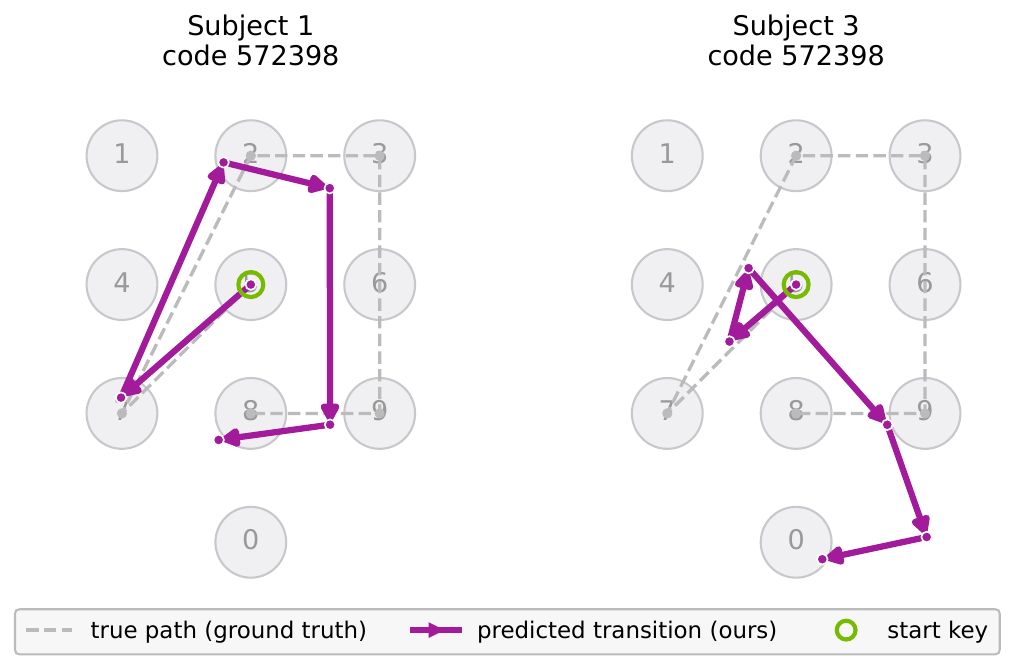}
    \caption{Estimated trajectories of the best case (Subject 1) and worst case (Subject 3)  when recovering a sample passcode.}
    \label{fig:gaze-trajectory}
\end{figure}

The central pattern is that stronger passcode recovery coincides with more clearly separated recovered keypad geometry. 
Figure~\ref{fig:calib-vs-passcode} visualizes recovered keypad geometry and helps explain the observed pattern. Subjects~1 and~2 produce separated landing clusters that preserve the keypad's spatial organization, whereas Subject~3's clusters are compressed toward the center and overlap. This pattern is consistent with lower externally observable head-motion amplitude for Subject~3, but does not necessarily establish that the subject relied primarily on eye rotation because eye motion was not measured independently. 

Figure~\ref{fig:gaze-trajectory} provides a complementary transition-level diagnostic for one example passcode for the best (Subject 1) and the worst case (Subject 3). For visualization, both trajectories are anchored at the ground-truth initial digit and formed by integrating the five continuous displacement estimates. Subject 1 preserves the approximate path geometry, whereas the first two predicted displacements of Subject 3 have notably smaller amplitudes than the corresponding true movements, after which the accumulated trajectory drifts away from the keypad path. This contraction could reflect Subject 3's greater eye--in--head contribution to those gaze shifts, but we cannot attribute it to eye compensation without an independent eye-motion measurement. Note that this diagnostic visualization shows only transition-level error and does not depict the exhaustive passcode-ranking procedure.

\subsection{Effect of Repeated Observations}
\label{subsec:repeat}

\begin{figure}[t]
    \centering
    \begin{subfigure}{0.49\linewidth}
        \includegraphics[width=\linewidth]{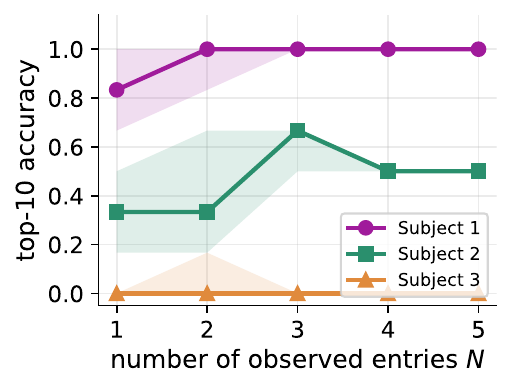}
        \caption{Top-10 accuracy}
    \end{subfigure}
    \hfill
    \begin{subfigure}{0.49\linewidth}
        \includegraphics[width=\linewidth]{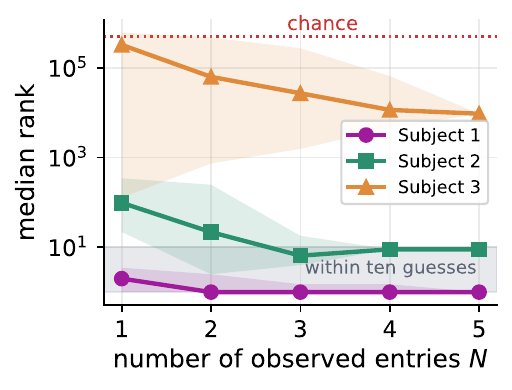}
        \caption{Median rank}
    \end{subfigure}
    \caption{Effect of the number of observed entries $N$ on (a) top-10 accuracy and (b) median rank. Lines show the median over all combinations of choosing $N$ entries from the five available recordings, and shaded bands span the corresponding minimum and maximum numbers.}
    \label{fig:repeat}
\end{figure}

Our threat model assumes that the adversary can record the same passcode more than once (Section~\ref{sec:threat}). To see if repeated observations help an adversary recover the true passcode, we recorded the test passcode five times. 
To avoid privileging an arbitrary recording order, we evaluate all $\binom{5}{N}$ subsets for each observation budget $N\in\{1,\ldots,5\}$. 
For a given subset, the selected entries are pooled position by position at decoding time; this operation does not retrain or otherwise change the transition scorer. Pooling combines only the five recordings of the same passcode. The other five test passcodes are not used, because a victim has only one device passcode. We evaluate each of the six test passcodes separately in this way. We then compute top-10 accuracy and median rank across the six test passcodes. Figure~\ref{fig:repeat} plots, for each subject and $N$, the median of each metric over the $\binom{5}{N}$ subsets, with the shaded band spanning the minimum and maximum. The bands therefore measure sensitivity to which recordings are available.

The benefit of repetition differs across subjects. For Subject~1, the median $N=1$ subset already achieves top-10 accuracy 1.00 and median rank 2, so additional observations provide little improvement.
Subject~2 benefits more: its median rank decreases from 174 at $N=1$ to 19 at $N=2$ and 9 at $N=4$, while top-10 accuracy increases from 0.17 to 0.50. 
Subject~3 does not reach the top-10 criterion, but its median rank decreases from 251,507 at $N=1$ to 9,528 when all five entries are pooled. 
Repeated observations can therefore reinforce consistent evidence and substantially reduce the search space even when they do not produce immediate top-10 recovery. The min--max bands also show that the gain is not determined by $N$ alone: different subsets at the same observation budget can produce substantially different rankings. For Subject~3, the best single recording yields a median true-passcode rank of 128, whereas the worst performs near the random-ranking baseline.

\noindent\textbf{Exploratory Entry Selection.}
The subset variability in Figure~\ref{fig:repeat} raises a separate question: can the adversary identify an informative recording without knowing the true passcode? To this end, let $E_r$ be the mean predicted motion magnitude in entry $r$.
We measure how much these values differ across the five entries with the
label-free coefficient of variation
$E_{\mathrm{cv}}=\operatorname{sd}(E_1,\ldots,E_5)/\operatorname{mean}(E_1,\ldots,E_5)$. We then consider two different entry selection strategies: (i) \emph{all entries}: an adversary averages all entries; (ii) \emph{best entry}: an adversary chooses the entry with the largest predicted head-motion magnitude. Note that selecting recordings with large predicted head-motion magnitude is a heuristic rather than a validated optimal strategy. Establishing its generalizability requires evaluation on a substantially larger cohort with the selection rule fixed in advance.

Table~\ref{tab:movement-adaptive} shows that $E_{\mathrm{cv}}$ predicts which
strategy wins. When the entries are alike ($E_{\mathrm{cv}}=0.10$ and $0.12$ for
Subjects~1 and~2), averaging cancels per-entry noise and pooling all five is
best; keeping a single entry instead carries its noise and can hurt sharply,
raising Subject~2's median rank from $9$ to $84$. When the entries differ
($E_{\mathrm{cv}}=0.49$ for Subject~3, whose quietest recording carries far less
predicted motion than its strongest), pooling dilutes the one informative entry,
so retaining the largest-motion entry is far better and cuts the median rank from
$9{,}528$ to $128$. A label-free rule that pools all entries at low
$E_{\mathrm{cv}}$ and keeps only the largest-motion entry at high
$E_{\mathrm{cv}}$ therefore matches the better fixed strategy for every subject
(last column of Table~\ref{tab:movement-adaptive}). Because the selection uses
only predicted displacement magnitudes, it requires no labels and remains
available in the cross-person setting. 

\begin{table}[t]
\centering
\caption{Exploratory comparison of entry-combination strategies under the within-person protocol.}
\label{tab:movement-adaptive}
\small
\begin{tabular}{l c c c c}
\toprule
& & \multicolumn{2}{c}{\textbf{Entry selection strategy}} & \\
\cmidrule(lr){3-4}
\textbf{Subject} & \textbf{$E_{cv}$} & \textbf{All entries} & \textbf{Best entry} & \textbf{Selected strategy} \\
\midrule
Subject 1 & 0.10 & \textbf{1}       & 2            & All entries \\
Subject 2 & 0.12 & \textbf{9}       & 84           & All entries \\
Subject 3 & 0.49 & 9{,}528          & \textbf{128} & Best entry  \\
\bottomrule
\end{tabular}
\end{table}

\subsection{Effect of Passcode Structure}
\label{subsec:codediff}

\begin{figure}[t]
    \centering
    \includegraphics[width=0.6\linewidth]{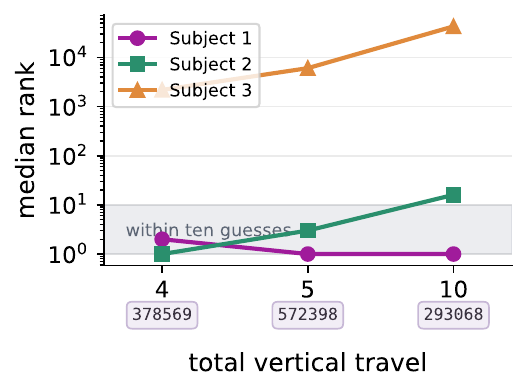}
    \caption{Passcode recovery for the three passcodes shared by all subjects under the within-person protocol. Each line shows one subject's median true-passcode rank as total vertical keypad travel increases from 4 to 10 steps.}
    \label{fig:codegeom}
\end{figure}

To examine whether keypad-path geometry affects recovery, we focus on the three passcodes shared by all subjects. This matched subset holds the passcode constant across subjects and avoids comparing different personal codes. For a passcode $D$, we summarize its path by total horizontal and vertical travel, $T_h=\sum_{j=0}^{4}|\Delta\mathrm{col}_j|$ and $T_v=\sum_{j=0}^{4}|\Delta\mathrm{row}_j|$, measured in keypad steps. The three shared codes have the same horizontal travel, $T_h=4$, but increasing vertical travel: $T_v=4$ for 378569, $T_v=5$ for 572398, and $T_v=10$ for 293068.

Figure~\ref{fig:codegeom} shows an association between greater vertical travel and poorer recovery. The subject-level pattern is not uniform, however. The ranks for Subjects~2 and~3 worsen as vertical travel increases, whereas Subject~1 remains at rank 1 or 2 for all three codes. This subject is already near the best possible rank, leaving little room for the code structure to produce a visible monotonic effect.
Unfortunately, our preliminary experiment is not designed to isolate passcode-structure effects. Figure~\ref{fig:codegeom} only provides exploratory evidence of passcode-dependent difficulty, not an estimate of the effect of vertical travel. A future study should use a larger, balanced set of passcodes that independently varies horizontal and vertical travel, step lengths, and direction changes with randomized passcode assignment.

\subsection{Robustness to Video Capture Quality}
\label{subsec:quality}

\begin{figure*}[!th]
    \centering
    \begin{subfigure}{0.24\linewidth}
        \includegraphics[width=\linewidth]{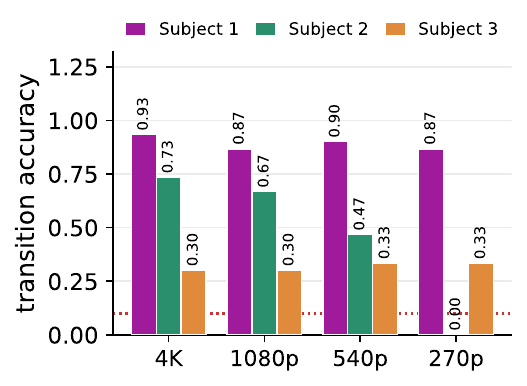}
        \caption{Transition acc. per resolution}
        \label{fig:qual_a}
    \end{subfigure}\hfill
    \begin{subfigure}{0.24\linewidth}
        \includegraphics[width=\linewidth]{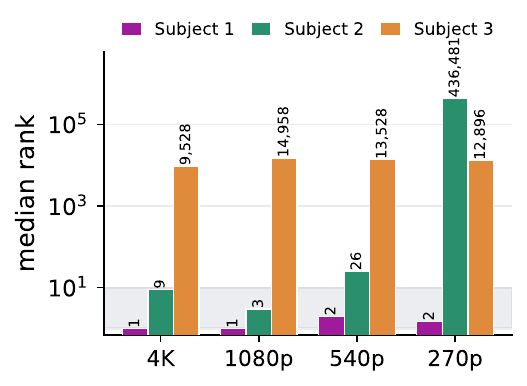}
        \caption{Med. rank per resolution}
        \label{fig:qual_b} 
    \end{subfigure}\hfill
    \begin{subfigure}{0.24\linewidth}
        \includegraphics[width=\linewidth]{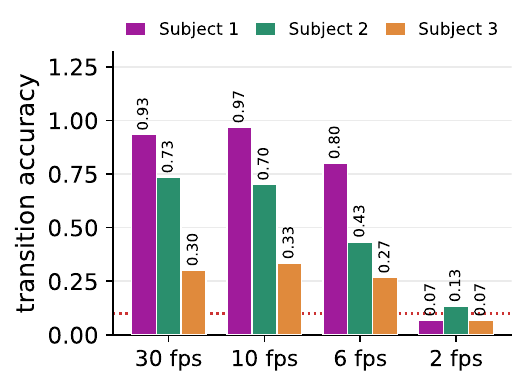}
        \caption{Transition acc. per FPS}
        \label{fig:qual_c} 
    \end{subfigure}\hfill
    \begin{subfigure}{0.24\linewidth}
        \includegraphics[width=\linewidth]{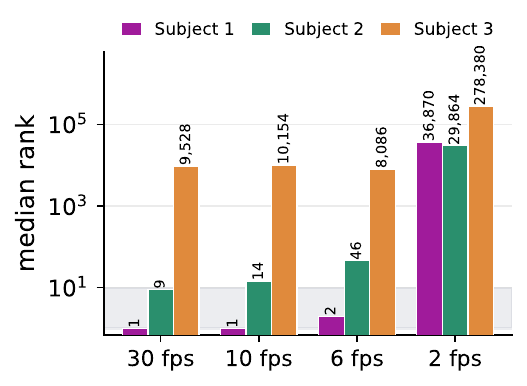}
        \caption{Med. rank per FPS}
        \label{fig:qual_d} 
    \end{subfigure}
    \caption{Effect of inference performance per captured video quality. (a, b) Transition accuracy and median rank across capture resolutions. (c, d) Transition accuracy and median rank across frame rates. Recovery remains effective down to 540p and 6 fps.}
    \label{fig:quality}
\end{figure*}

The recordings used in our main experiments (Section~\ref{sec:inference-results}) rely on 4K resolution at 30 frames per second, a setting
available on commodity smartphones. In a surveillance setting, however, the
wearer's head may occupy only a small part of a wide-field recording, so
cropping the head produces a substantially lower-resolution input to the
tracker. Stored CCTV footage may also have a lower frame rate~\cite{surveillancecommissioner2018}.
These conditions motivate evaluating both
spatial and temporal degradation.
To isolate the effect of each sampling axis, we degrade the same source videos
along one axis at a time and rerun the pipeline.

\noindent\textbf{Varying Video Resolution.} Figure~\ref{fig:qual_a} and \ref{fig:qual_b} show the transition accuracy and median ranks when varying the video resolutions. As video resolution gets lower, we observe that the inference performance is preserved to 540p, but degrades at 270p where the head region spans roughly one hundred and sixty pixels. Considering that 270p is an extreme condition, our attack is robust at standard video resolutions. Additionally, the degradation is not uniform across subjects. Subject 2 falls from a transition accuracy of 0.47 to 0.00, and its median rank rises from 26 to 436,481, whereas Subject 1 still reaches 0.87 and Subject 3 is already low enough that little changes.

\noindent\textbf{Varying Video Frame Rates.} Figure~\ref{fig:qual_c} and \ref{fig:qual_d} show the transition accuracy and median ranks when varying the video frame rates. As our pipeline relies on 30 fps, we evaluate lower frame rates, where temporal information is progressively lost. Inference performance remains stable at most lower frame rates before failing abruptly. At 6 fps, the transition accuracy remains 0.80 for Subject 1 and 0.43 for Subject 2, with no change in the order of magnitude of the median rank. At 2 fps, however, all three subjects fall to a transition accuracy between 0.07 and 0.13, effectively reaching chance performance. This suggests that the descriptor remains robust to moderate temporal degradation but fails once too little temporal information is retained. In contrast, reducing the spatial resolution degrades performance more steadily, with failure occurring at 270p.

Overall, these findings indicate that the attack does not require specialized capture equipment. A standard surveillance camera or smartphone operating at 1080p provides sufficient detail, supporting the practical viability of the threat model described in Section~\ref{sec:threat}.

\subsection{Pipeline Ablations}
\label{sec:ablation}
We now justify the main design choices in \ourtool{} through a sequence of
ablations. We (i)~replace the whole descriptor-and-readout stack with a frozen
learned end model, (ii)~vary the transition label granularity and the readout while holding the rest of the pipeline fixed, and (iii)~vary the descriptor subset that the scorer consumes. Unless stated otherwise, all numbers are reported
within-person and averaged over the three subjects

\noindent\textbf{Pretrained Visual Representations.} We first replace our motion descriptor and regressor with a frozen pretrained model that predicts the destination key of each transition through a linear classifier trained on its frozen features. The image models receive the final frame of each
transition, whereas the video models receive the complete transition clip.

\begin{table}[htb]
\centering
\caption{Evaluating frozen learned models as replacements for our pipeline. Image and video models take the final frame and the full clip as input, respectively.}
\label{tab:endmodel}
\small
\resizebox{\columnwidth}{!}{
\begin{tabular}{l l c c c}
\toprule
& & \textbf{Transition} & \multicolumn{2}{c}{\textbf{Passcode}} \\
\cmidrule(lr){3-3}\cmidrule(lr){4-5}
\textbf{Model} & \textbf{Temporal Input} & \textbf{Acc.} & \textbf{Top-10 Acc.} & \textbf{Med. Rank} \\
\midrule
VideoMAE-B     & Transition clip        & 0.36 & 0.11 & 12{,}526 \\
V-JEPA2 ViT-L  & Transition clip        & 0.33 & 0.00 & 14{,}536 \\
DINOv2 ViT-L   & Final transition frame  & 0.26 & 0.00 & 101{,}520 \\
R(2+1)D-18     & Transition clip        & 0.20 & 0.00 & 161{,}066 \\
CLIP ViT-L     & Final transition frame  & 0.16 & 0.00 & 393{,}790 \\
\midrule
\textbf{Ours (LK)} & \textbf{Transition clip} & \textbf{0.66} & \textbf{0.50} & \textbf{9} \\
\bottomrule
\end{tabular}
}
\end{table}
Table~\ref{tab:endmodel} compares VideoMAE~\cite{tong2022videomae}, V-JEPA~2~\cite{assran2025v}, DINOv2~\cite{oquab2023dinov2}, R(2+1)D~\cite{tran2018closer}, and CLIP~\cite{radford2021learning}. The two strongest learned alternatives are the video models: VideoMAE reaches 0.36 transition accuracy and V-JEPA~2 reaches 0.33. However, R(2+1)D performs below the image-based DINOv2 model, indicating that temporal input alone does not explain performance. The best learned alternative, VideoMAE, yields a median true-passcode rank of 12,526, compared with 9 for \ourtool{}. This experiment establishes a substantial performance gap for the tested configurations, but it does not by itself identify the cause; Section~\ref{sec:discuss:modern} discusses reduced feature resolution as one plausible explanation.

\begin{table}[htb]
\centering
\caption{Joint effect of transition-label granularity and readout under the
within-person protocol with all five entries pooled. }
\label{tab:granularity}
\small
\resizebox{\columnwidth}{!}{%
\begin{tabular}{c l c c}
\toprule
\textbf{Granularity} & \textbf{Readout} &
\textbf{Top-10 Acc.} & \textbf{Median Rank} \\
\midrule
                        & nearest centroid        & 0.22 & 4{,}086 \\
Exact-pair $\pi$ & multinomial classifier  & 0.17 & 1{,}751 \\
                        & ridge regression         & \multicolumn{2}{c}{N/A} \\
\midrule
                       & nearest centroid        & 0.17 & 846 \\
\textbf{Movement $\mu$}   & multinomial classifier  & 0.28 & 356 \\
                       & \textbf{ridge regression (ours)} & \textbf{0.50} & \textbf{9} \\
\midrule
                       & nearest centroid        & 0.06 & 3{,}730 \\
Direction $\delta$ & multinomial classifier & 0.22 & 165 \\
                       & ridge regression         & 0.11 & 258 \\
\bottomrule
\end{tabular}%
}
\end{table}

\noindent\textbf{Label Granularity and Readout.} 
Table~\ref{tab:granularity} crosses three transition-label spaces (Section~\ref{sec:transition-model}) with the
applicable readouts. Exact-pair labels $\pi$ retain the source and destination
digits as 100 categorical labels. Movement labels $\mu$ merge transitions with
the same two-dimensional keypad displacement into 31 labels, and direction
labels $\delta$ retain only the signs of the horizontal and vertical
displacements, producing 9 labels. The nearest-centroid and multinomial
classifier readouts apply to all three spaces. Ridge regression applies to the
vector-valued movement and direction spaces, but not to arbitrary exact-pair
identifiers, which have no meaningful Euclidean ordering.

Among the three label granularities, movement $\mu$ gives the best passcode recovery, and it matches the mechanics of gaze-coupled head motion, which depends chiefly on how far the gaze travels rather than on the absolute source and destination keys. Because displacement lives in a continuous, ordered space, we read it out with a ridge regressor rather than a classifier, which improves the median rank from $356$ to $9$; a regression readout is either inapplicable to the unordered exact-pair or inferior in the case of the most coarse granularity of direction.

\begin{table}[t]
\centering
\caption{Descriptor-subset selection comparison.}
\label{tab:featsel}
\small
\resizebox{\columnwidth}{!}{
\begin{tabular}{c c c c c}
\toprule
& & \textbf{Transition} & \multicolumn{2}{c}{\textbf{Passcode}} \\
\cmidrule(lr){3-3}\cmidrule(lr){4-5}
\textbf{\#~dims} & \textbf{Selection} & \textbf{Acc.} & \textbf{Top-10 Acc.} & \textbf{Median Rank} \\
\midrule
100 & All entries & 0.48 & 0.33 & 635 \\
20  & PCA         & 0.59 & 0.50 & 26  \\
\midrule
14  & Random      & 0.43 & 0.28 & 618 \\
14  & PCA         & 0.62 & 0.50 & 12  \\
\textbf{14} & \textbf{Custom (ours)} & \textbf{0.66} & \textbf{0.50} & \textbf{9} \\
\midrule
8   & PCA         & 0.59 & 0.44 & 23  \\
\bottomrule
\end{tabular}
}
\end{table}

\noindent\textbf{Descriptor Subset.}
\label{sec:subset}
Table~\ref{tab:featsel} reports the inference performance with selection strategies of features that are fed into the descriptor (Section~\ref{subsec:description}).
Note that our $100$-dimensional descriptor is deliberately over-complete: it collects many summary statistics per motion channel so that no informative cue is missed at extraction time. Feeding all $100$ entries to the scorer, however, is counter-productive---each subject supplies only several hundred reference transitions, which cannot constrain a $100$-dimensional ridge map, and the resulting overfit inflates the median rank to $635$. The scorer therefore consumes a reduced subset, and two decisions make that subset principled rather than arbitrary: how many entries to keep, and which ones.

Comparing the PCA (Principal Component Analysis) baseline, the median rank peaks at $14$ (rank $12$) and is worse at both $8$ (rank $23$) and $20$
(rank $26$). Thus, $14$-dimension is the empirical knee that balances descriptive capacity against the variance budget of a several-hundred-transition reference set.
At a fixed dimension, \emph{which} entries are kept matters more than the count: a random $14$-entry subset reaches only median rank $618$, because most descriptor statistics are redundant or weakly coupled to keypad displacement. We instead retain the entries that carry the displacement signal most directly, i.e., the summary statistics of the translation and in-plane-rotation channels together with the path-geometry terms.
Unlike a PCA projection, which is competitive (rank $12$ at $14$ dimensions) but mixes all channels into opaque components, this subset preserves the physical reading of every axis, which is what lets the recovered displacements be interpreted as keypad geometry (Section~\ref{sec:inference-results}). 
The subset is chosen once and applied unchanged to every subject and to both protocols.

%% file: sections/discussion.tex
\section{Discussion}
\label{sec:discussion}
This section interprets the residual ranking beyond the top-10 criterion (Section~\ref{sec:discuss:search-space-impact}), explains why modern visual models underperform our classical pipeline (Section~\ref{sec:discuss:modern}), and then discusses a behavioral confound, countermeasures, limitations, and ethical considerations.

\subsection{Security Impact Beyond Top-10 Recovery}
\label{sec:discuss:search-space-impact}
Top-10 recovery captures an immediately actionable online attack under the guess
budget used in our evaluation, but a top-10 miss does not imply an absence of
leakage.  For Subject~3, the median true-passcode rank is 9,528 within-person and
12,786 cross-person (Section~\ref{sec:inference-results}).  Thus, at the median, the true passcode lies within the top
0.95\% and 1.28\% of the $10^6$-code space, respectively, rather than near the
median rank of 500,000. This reduction is insufficient to unlock an
unmodified device within ten attempts, but it shows that Subject~3 still leaks
information that can materially prioritize a larger guessing campaign.

The operational significance of this reduction depends on a second capability:
the adversary must obtain more verification attempts than the normal interface
permits.  Prior work such as bypassing the retry counter with NAND mirroring~\cite{skorobogatov2016bumpy} and parallelized brute-forcing with UID key extraction~\cite{lisovets2021let} have demonstrated such capabilities on earlier Apple devices. 
They do not, however, demonstrate such a compromise against Apple Vision
Pro.  Apple's current design binds passcode derivation to a per-device UID and
enforces delays and attempt counters through the Secure Enclave~\cite{apple_secure_enclave}.  
Our claim is therefore conditional: if a well-resourced adversary can
separately obtain an offline verifier, reset verification state, or instantiate
independent verifier copies, \ourtool{}'s ranked output substantially reduces the work
of the ensuing search.

The residual rank also identifies opportunities to strengthen the inference stage.  
Training on a larger and more diverse participant cohort could better cover variation in eye--head coordination and reduce cross-person domain shift.  
Multiple synchronized viewpoints could provide complementary projections of the same small rotation and remain informative when one view is occluded or nearly insensitive to a motion axis.  
Repeated observations of the same passcode could likewise be pooled, with motion-aware weighting or robust aggregation so that weak entries do not dilute stronger ones.

\subsection{Why Modern Visual Models Fall Short}
\label{sec:discuss:modern}
\ourtool{} utilizes the classical vision-based tracking method to quantify the user's head movement (Section~\ref{sec:attack}) instead of utilizing recent advanced vision or video models. 
Our design choice is based on our benchmark results (see Table~\ref{tab:endmodel}) showing that frozen learned representations underperform with a significant margin. Even the strongest learned alternative, VideoMAE-B, records several orders of magnitude higher median rank.

A plausible explanation is a mismatch between the representations and the signal of interest.
\ourtool{} must distinguish sub-degree head rotations that may produce only subpixel or few-pixel displacements in the captured video.
Modern image and video encoders commonly reduce spatial resolution through strided feature maps~\cite{tran2018closer}, patch tokens~\cite{oquab2023dinov2,radford2021learning}, or spatio-temporal tubelets~\cite{tong2022videomae,assran2025v}.
Although these representations support strong semantic and long-range correspondence performance, they may attenuate the subpixel-to-few-pixel displacement that carries our side-channel signal. 

\subsection{Prompt Consultation as a Behavioral Confound}
The performance difference across subjects may partly reflect how they performed the experimental task, rather than only an inherent difference in eye--head coordination.  During post hoc review, Subject~3, who is also an author, reported repeatedly consulting the assigned passcode displayed above the keypad (Figure~\ref{fig:keypad_image}, right) instead of memorizing the complete six-digit sequence before entry.  
Such prompt consultation violates an assumption of our transition model: the motion between two consecutive pinch events may include a gaze shift toward the prompt and back, rather than primarily the direct movement from one keypad digit to the next.  The resulting composite trajectory can distort the direction and amplitude assigned to a digit transition and may have contributed to Subject~3's lower accuracy.  
This explanation is retrospective, however, and cannot be established from the present data because prompt-directed glances were not annotated or experimentally controlled.

This behavior differs from our intended device-unlock scenario, in which a user normally enters a familiar passcode without consulting an on-screen copy. Future studies should therefore separate passcode familiarity from inference difficulty.  Before each entry, participants should be shown an assigned synthetic passcode, required to memorize all six digits, and allowed to begin only after the prompt has been removed and recall has been verified.  To better approximate habitual entry without collecting real credentials, participants could practice synthetic passcodes to a predefined accuracy criterion before recording.  A controlled comparison among visible-prompt, newly memorized, and practiced-passcode conditions would then quantify whether prompt consultation explains part of the per-subject performance gap.  Until such an experiment is performed, Subject~3's result should not be attributed solely to a stable low-head-motion profile.

\subsection{Countermeasure}
\noindent\textbf{Physical Leakage Beyond Software Control.} 
Unlike previous keystroke or gaze-inference attacks that can be prevented by a software patch (Section~\ref{sec:related}), \ourtool{} exploits a biomechanical signal---the head reorientation that
the central motor program couples to every gaze shift
(Section~\ref{sec:bg-eye-head})---from an external camera the device does not
control. A vendor cannot completely suppress it without redesigning gaze interaction itself.

\noindent\textbf{Mitigations.}  
One direct defense is to shrink the spatial extent of sensitive UI elements. If all passcode buttons lie within a small angular region near the current fixation, gaze shifts can be completed mostly by eye rotation and are less likely to recruit measurable head motion. This \textit{micro-UI} mode could be reserved for short secrets such as passcodes or OTP entry, rather than used throughout the interface. The cost is usability: smaller targets increase visual crowding, selection errors, dwell time, and accessibility burden, especially for users with reduced acuity, tremor, or less stable eye tracking. A secure design would therefore need to quantify the privacy gain against error rate and entry time.

Another simple yet effective defense is to randomize the keypad layout during sensitive input. Randomization weakens the attack because a fixed head-motion direction no longer maps consistently to the same digit transition. It is particularly useful against an adversary who pools multiple recordings of the same passcode: if the layout changes across entries, repeated observations do not reinforce the same ordered-pair classes. The drawback is again interaction cost. Randomized keypads slow entry, increase cognitive load, and make motor memory unusable. They may also increase legitimate lockouts if users enter familiar passcodes under an unfamiliar layout. A practical system might therefore use partial randomization, row/column shuffling, or randomization only after risk triggers such as public-location use, recent failed attempts, or
external-camera warnings. We view robust mitigation as an open problem and a motivation for future research.

\subsection{Limitation and Future Direction}
\label{sec:limits}
There are several limitations. First, the current data
come from author subjects rather than recruited unaware participants. 
The knowledge of attack hypothesis can bias head behavior in either direction: a subject may consciously hold the head still and rely on eye movement alone --- suppressing the very signal we measure --- or may over-rotate the head toward each key, inflating it. Either way, the head motion recorded here is unlikely to match that of an unaware user entering a passcode out of habit. The experimental protocol for recruited participants should be carefully designed to differentiate the effect of the subject's awareness of attack hypothesis that can be expressed by suppressing their head motion. 
Second, the front perspective is favorable because the face and headset are visible; real
surveillance may observe the user from the side, behind, or under partial
occlusion. Third, the six-digit passcode setting is only one gaze-input task. 
Passwords, OTPs, text entry, app focus, and document selection may have different leakage profiles. Finally, we only evaluate our attack against a single device, i.e., Apple Vision Pro. 

We plan to address these limitations with an IRB-approved user study,
more participants, multiple camera perspectives, carefully designed experiment protocol and additional tasks beyond numeric passcodes. 

\begin{table*}[t]
\centering
\caption{Qualitative comparison of \ourtool{} with prior keystroke and gaze-inference
attacks on XR devices. ``External observ.'' indicates that the exploited signal
can be captured by sensing the victim from outside the headset; ``No SW install''
indicates that the attack requires no attacker-controlled software on the
victim's device; and ``Resists SW patch'' indicates that the leakage comes from
an externally measurable physical signal rather than a software-exposed data or
rendering channel.}
\label{tab:comparison}
\small
\setlength{\tabcolsep}{4pt}
\begin{tabular}{l l l c c c}
\toprule
\textbf{Attack} & \textbf{Exploited signal} &
\textbf{Input modality} &
\makecell{\textbf{External}\\\textbf{observ.}} &
\makecell{\textbf{No SW}\\\textbf{install}} &
\makecell{\textbf{Resists SW}\\\textbf{patch}} \\
\midrule
GAZEploit~\cite{wang2024gazeploit}      & Avatar eye animation         & Gaze-as-pointer    & \xmark & \cmark & \xmark \\
TyPose~\cite{slocum2023going}          & Head motion (IMU)            & Controller/hand & \xmark & \xmark & \xmark \\
Hidden Reality~\cite{gopal2023hidden}   & Hand-gesture video           & Controller/hand & \cmark & \cmark & \cmark \\
VR-Spy~\cite{al2021vr}           & Wi-Fi CSI                    & Controller/hand & \cmark & \cmark &  \cmark \\
VRecKey~\cite{ni2024non}              & Controller IR LEDs           & Controller/hand & \cmark & \cmark & \cmark \\
Yang et~al.~\cite{yang2024can}  & Avatar hand motion           & Phys.\ keyboard & \xmark & \cmark & \xmark \\
\midrule
\textbf{\ourtool{} (ours)}              & \textbf{Head motion (video)} & \textbf{Gaze-as-pointer} & \cmark & \cmark & \cmark \\
\bottomrule
\end{tabular}
\end{table*}

\subsection{Ethical Concerns}
\label{sec:ethics}
All recordings were collected from the authors themselves. Because the
study involves no third-party participants, it does not constitute human-subjects
research and does not require institutional review board (IRB) approval. Nor are
any real secrets exposed: the entered codes are experimental stimuli generated
for the study, not genuine device passcodes or account credentials, so neither
our data nor its release reveals a real passcode. The larger
study with recruited, unaware participants that we outline as future work will be
conducted only under IRB approval, with informed consent and no collection of
participants' real credentials.

Although we evaluate \ourtool{} on Apple Vision Pro, the observed leakage should not be interpreted as a defect specific to the vendor or device. It arises from gaze-coupled head motion observable outside the headset and may therefore affect other systems that use gaze to select spatially separated interface elements. Nevertheless, because Apple Vision Pro is our evaluated platform and these findings may inform the design of its security-sensitive interfaces, we have shared our preliminary findings with Apple by filing a security report. We also present this issue as a broader design-level concern for the gaze-as-pointer research community.

%% file: sections/related.tex
\section{Related Work}
\label{sec:related}

The privacy implications of eye-tracking sensors integrated into consumer XR headsets have only recently begun to attract systematic attention. Most closely related to our work is GAZEploit~\cite{wang2024gazeploit}, which infers the gaze direction of an Apple Vision Pro user by analyzing the eye animation of the user's Persona avatar broadcast through a video call. While GAZEploit and \ourtool{} share the underlying goal of recovering gaze without direct access to the device's eye-tracker, the two attacks differ fundamentally in threat model: GAZEploit assumes the victim is actively sharing their Persona via a video stream while \ourtool{} operates against any wearer visible to an external camera, regardless of avatar broadcast or OS version, and exploits a \emph{biomechanical} leak (eye--head coordination) that vendors cannot fully eliminate without redesigning the gaze-based interaction.

On VR headsets that predate gaze-as-pointer interaction, TyPose~\cite{slocum2023going} showed that virtual-keyboard input on VR headsets can be reconstructed from the wearer's \emph{head-pose telemetry} alone. The attack relies on the same biomechanical signal we exploit --- head motion synchronized with gaze --- but operates from a fundamentally different vantage point: it assumes the attacker has installed a \emph{background application on the headset itself}, where Inertial Measurement Unit (IMU) data are accessible without any special permission. \ourtool{} removes this assumption entirely: head motion is observed \emph{externally}, from a commodity camera, with no software running on the victim's device. Table~\ref{tab:comparison} summarizes related work. \ourtool{} is the first inference attack against gaze-as-pointer XR devices from local and external observation by an adversary.

A second line of work studies the use of eye-tracking data itself
as a biometric signal. Aziz and Komogortsev~\cite{aziz2025exploring}
show that gaze streams exposed to third-party applications --- in
combination with head and hand motion streams --- enable
re-identification of XR users with high accuracy even when
state-of-the-art privacy mechanisms are applied to each stream
independently. These attacks differ from \ourtool{} in that they
\emph{require} the gaze stream to be available to the attacker (i.e.,
they target the \emph{output} of the eye tracker), whereas \ourtool{}
reconstructs gaze direction without observing it directly.

%% file: sections/ack.tex
\section*{Acknowledgements}
This work was partially supported by Institute of Information \& communications Technology Planning \& Evaluation (IITP) [RS-2023-00215700, Trustworthy Metaverse: blockchain-enabled convergence research, 70\%] and the National Research Foundation of Korea (NRF) grant funded by the Korea government (MSIT) [RS-2026-25495427, 30\%].

%% file: sections/conclusion.tex
\section{Conclusion}
\label{sec:conclusion}
We presented \ourtool{}, a preliminary external-observation attack on gaze-based passcode entry in XR. The attack exploits gaze-coupled head motion that can be observed by a nearby adversary without access to the display or eye-tracking data. Although preliminary, our evaluation suggests that protecting gaze input may require more than concealing visual output and eye-tracking streams; it may also require mitigating information leakage through users' physical movements. We hope this work motivates the development of practical defenses and further investigation of gaze-input privacy under a broader range of users, adversaries, and real-world attack scenarios.